\documentclass[aps,11pt,axodraw,nofootinbib]{revtex4}
\pdfoutput=1
\usepackage{epsfig}
\usepackage{amsmath}
\usepackage{bm}
\usepackage{times}
\usepackage{graphicx}
\usepackage{epstopdf}
\usepackage{amsfonts}
\usepackage{bm}
\usepackage{epsfig}
\usepackage{graphics}
\usepackage{xspace}
\usepackage[usenames]{color}

\def\nn{\nonumber}
\def\bea{\begin{eqnarray}}
\def\eea{\end{eqnarray}}
\def\ba{\begin{eqnarray}}
\def\ea{\end{eqnarray}}
\def\be{\begin{equation}}
\def\ee{\end{equation}}

\def\beq{\begin{equation}}
\def\eeq{\end{equation}}

\def\nn{\nonumber}

\begin{document}
\title{\Large An expansion for Neutrino Phenomenology}
\author{ Benjam\'{i}n Grinstein$^{1}$}
\email{bgrinstein@ucsd.edu}
\author{Michael Trott$^{2}$}
\email{michael.trott@cern.ch}

\affiliation{$^{1}$ Department of Physics, University of California, San Diego, La Jolla, CA 92093 USA}

\affiliation{$^{2}$Theory Division, Physics Department, CERN, CH-1211 Geneva 23, Switzerland}

\date{\today}
\begin{abstract}
We develop a formalism for constructing the Pontecorvo-Maki-Nakagawa-Sakata (PMNS) matrix and neutrino masses using an expansion that originates when a sequence of heavy right handed neutrinos are integrated 
out, assuming a seesaw mechanism for the origin of neutrino masses. 
The expansion establishes relationships between the structure of the PMNS matrix and the mass differences of neutrinos,
and allows symmetry implications for measured deviations from tri-bimaximal form to be studied systematically.
Our approach does not depend on choosing the rotation between the weak and mass eigenstates
of the charged lepton fields to be diagonal. We comment on using this expansion to examine the symmetry implications of the recent results
from the Daya-Bay collaboration reporting the discovery of a non zero value for $\theta_{13}$, indicating a deviation from tri-bimaximal form, with a significance of $5.2 \, \sigma$. 
\end{abstract}
\maketitle
\section{Introduction}
The standard theory of neutrino oscillations, where three mass eigenstate neutrinos differ from their interaction eigenstates
leading to the observed neutrino oscillations, is consistent with current experimental data.
The amplitude of the oscillations among various neutrino species is related to the misalignment of the
interaction and mass eigenstates of the neutrinos, characterized by the Pontecorvo-Maki-Nakagawa-Sakata (PMNS) matrix \cite{Pontecorvo:1957cp,Maki:1962mu}. 
Taking the relation between the interaction (primed) and mass (unprimed) eigenstates to be given by
$f'_{L/R} = \mathcal{U}(f,L/R) \, f_{L/R}$, where $f = \{\nu,e,\mu, \cdots \}$, the PMNS matrix is given by
\bea
\mathcal{U}_{PMNS} = \mathcal{U}^\dagger(e,L) \,  \mathcal{U}(\nu,L).
\eea
The $\mathcal{U}_{PMNS}$ matrix can be parameterized in terms of three angles $\theta_{12},\theta_{13},\theta_{23}$ and three CP violating phases
$\delta, \alpha_{1,2}$. Defining $s_{ij} = \sin \theta_{ij}$ and $c_{ij} = \cos \theta_{ij}$ with the conventions $0 \le \theta_{ij} \le \pi/2$,
and $0 \le \delta, \alpha_{1,2} \le 2 \pi$ a general parameterization of this matrix is given by
\bea\label{param}
\mathcal{U}_{PMNS} &=& \left( \begin{array}{ccc}
1 & 0 & 0 \\
0 &  c_{23} & s_{23} \\
0 & -s_{23}& c_{23}  \end{array} \right) \times \left( \begin{array}{ccc}
c_{13} & 0 & s_{13} e^{- i \delta} \\
0 &  1 & 0 \\
- s_{13} e^{- i \delta}  & 0 & c_{13}  \end{array} \right) 
\times \left( \begin{array}{ccc}
c_{12} & s_{12} & 0 \\
-s_{12} &  c_{12} & 0 \\
0 & 0 &1  \end{array} \right) P,
\eea
where $P = {\rm diag}(e^{i \alpha_1/2},e^{i \alpha_2/2},1)$ is a function of the Majorana phases, present if the right handed neutrinos are
Majorana, while $\delta$ is a Dirac phase. This latter phase can contribute in principle to neutrino oscillation measurements, while the Majorana phases cannot.
Recent  global fit results on neutrino mass differences and measured mixing angles using old/new reactor fluxes
are given in Ref. \cite{Fogli:2011qn}, (with new reactor flux results in brackets):
\bea
\Delta m_{21}^2 &=& (7.58_{-0.26}^{+0.22}) \times 10^{-5} {\rm eV^2}, \nn \\
|\Delta \bar{m}_{32}^2| &=& (2.35_{-0.09}^{+0.12}) \times 10^{-3} {\rm eV^2}, \nn \\
\sin^2(\theta_{12}) &=& 0.306_{-0.15}^{+0.18} \, \, \quad (0.312_{-0.16}^{+0.17}), \nn \\
\sin^2(\theta_{13}) &=& 0.021_{-0.008}^{+0.007}  \quad (0.025_{-0.007}^{+0.007} ), \nn \\
\sin^2(\theta_{23}) &=& 0.42_{-0.03}^{+0.08}.
\eea
The error is the reported $1 \sigma$ error. Note that $ \Delta \bar{m}_{32}^2 \equiv m_3^2 - (m_1^2 + m_2^2)/2$ and
$\Delta \bar{m}_{32}^2 > 0, (< 0)$ corresponds to a normal (inverted) mass spectrum.
This pattern of experimental data is perhaps suggestive of a PMNS matrix that has at 
least an approximate ``tri-bimaximal''  $(\mathcal TB)$ form \cite{Harrison:2002er}. Fixing $\sin^2 (\theta_{12}) = 1/3$ and $\theta_{13} = 0$ the $\mathcal TB$ form is 
\bea
\mathcal{U}_{PMNS} \approx \mathcal{U}_{TB} = \left( \begin{array}{ccc}
\sqrt{\frac{2}{3}} & \frac{1}{\sqrt{3}} & 0 \\
-\frac{1}{\sqrt{6}} &  \frac{1}{\sqrt{3}}  & - \frac{1}{\sqrt{2}}  \\
-\frac{1}{\sqrt{6}} &  \frac{1}{\sqrt{3}}  & \frac{1}{\sqrt{2}}  \end{array} \right),
\eea
for a particular phase convention. 

The structure of this matrix could be fixed by underlying symmetries.
In attempting to determine such an origin of this matrix, the ``flavour'' basis
where one assumes $\mathcal{U}(e,L) = \rm{diag}(1,1,1)$ is frequently used. When this assumption is employed
the relationship between the weak and mass neutrino eigenstates is identified with the $\mathcal{U}_{PMNS}$, i.e.
$ \nu'_i = (\mathcal{U}_{PMNS})_{ij} \, \nu_j$ for $i,j = 1,2,3$.
There have been many attempts to link the approximate $\mathcal TB$ form of the neutrino mass matrix
to symmetries of the right handed neutrino interactions in this basis, see Ref. \cite{Altarelli:2004za,Altarelli:2010gt} for a review.
Recent experimental results provide evidence for deviations from this $\mathcal TB$ form. Evidence for $\sin^2(\theta_{13}) \neq 0$ in global fits is reported to be $> 3 \sigma$ in Ref. \cite{Fogli:2011qn} at this time.
As this paper was approaching completion, the discovery of non-vanishing $\theta_{13}$
was announced by the Daya Bay Collaboration \cite{An:2012eh} with a reported value of 
\bea
\sin^2 (2 \, \theta_{13}) = 0.092 \pm 0.016({\rm stat}) \pm 0.005({\rm syst}),
\eea
corresponding to $5.2 \sigma$ evidence for non zero $\theta_{13}$.
It is reasonable to expect further speculation about the origin of the deviation from
$\mathcal TB$ form in light of this result, where again the flavour basis will be frequently assumed. 

There is no clear experimental support for assuming that $\mathcal{U}(e,L) = \rm{diag}(1,1,1)$. This choice can be motivated
by an ansatz related to the origin of the approximately diagonal structure of the Cabibbo-Kobayashi-Maskawa (CKM) matrix, and a further ansatz on the relation between 
$\mathcal{U}^\dagger(d,L)$ and $\mathcal{U}^\dagger(e,L)$, see Ref. \cite{Altarelli:2004za} for a coherent discussion on this approach.
This choice can also be justified with model building in principle, see Ref. \cite{Csaki:2008qq} for an example.
Conversely, in grand unified models frequently associated with the high scale involved in the seesaw mechanism, the quark and lepton mass matricies 
can be related in such a manner that the flavour basis cannot be chosen, or at least,  the choice of the flavour basis can be highly artificial.

It is of interest to have a formalism for neutrino phenomenology that is as robust and basis independent as possible.
Clearly linking symmetries to the form of the PMNS matrix requires that the physical consequences of a symmetry 
are not dependent on a basis that can be arbitrarily choosen for $\mathcal{U}(e,L)$, such as the flavour basis.
In this paper, we develop a perturbative approach to the structure of the PMNS matrix
as an alternative to symmetry studies that are basis dependent.\footnote{This is a modern implementation in the neutrino sector of an old idea of relating the mixing matricies of the standard model (SM) to the measured quark or neutrino masses, for pioneering studies with this aim see Ref. \cite{Weinberg:1977hb,Fritzsch:1977za,Harari:1987ex,Froggatt:1978nt}.} 
Using this approach we show a basis independent relationship between the eigenvectors that give $\mathcal{U}(\nu,L)$ and $\mathcal{U}(e,L)$ corresponding to $\mathcal TB$ form at leading order in our expansion.
We also examine the prospects of relating patterns in the data in low scale measurements of neutrino properties with high scale flavour symmetry breaking, illustrating how our approach can be used to study
the recent reported discovery of non zero $\theta_{13}$ reported in Ref. \cite{An:2012eh}.

\section{Constructing a Flavour Space Expansion for Neutrino Phenomenology}
In this section we develop a formalism linking the measured
differences in the neutrino mass eigenstates with the structure of the
PMNS matrix, which we will refer to as a flavour space expansion
(FSE). There is an experimental ambiguity in the measured mass
hierarchies at this time. The neutrino mass spectrum can be a normal
hierarchy ($m_{3} > m_{2} \gtrsim m_1$) or an inverted hierarchy
($m_{3} < m_1 \lesssim m_2$).  In our initial discussion we will
assume a normal hierarchy. The formalism can be reinterpreted
for an inverted hierarchy.

\subsection{Review of Seesaw the Mechanism}
\label{seesaw}
Recall the standard seesaw scenario \cite{seesaw}, with three right handed neutrinos
$N_{Ri}$.\footnote{Our initial discussion will largely follow Ref.
  \cite{Broncano:2002rw,Broncano:2003fq,Jenkins:2008rb}.}  We use the
notation $\ell = (\nu_L,e_L)$ for the left handed $\rm SU(2)$ doublet
field and $e_R$ for the $\rm SU(2)$ singlet lepton field carrying
hypercharge.  These fields carry flavour indicies $i,j$.\footnote{These flavour indices will run over $a,b,c$ for the three $N'_i$. We use this notation
to distinguish these flavours from the measured mass differences of the physical eigenstates ($m_{1,2,3}$) and also the index on the Yukawa vectors
of each $N'_i$, which run over $1,2,3$.}  We also
define $\tilde{H} = i \, \tau_2 \, H^\star$, where $H$ is the
Higgs doublet of the SM, with $\langle H^T \rangle =
(0,v/\sqrt{2})$.  Then the lepton sector of the Lagrangian in the
seesaw scenario is the following 
\bea\label{seesaw} \mathcal{L} = i \,
\bar{N}'_{Ri} \, \partial \! \! \!  / \, N'_{Ri} - \frac{1}{2} \,
\overline{N^{\prime c}_{R \,i}} \, M'_{ij} \, N'_{R j} - \bar{\ell}'_{Li} \,
H \, (y'_E)_{ij} \, e'_{Rj} - \bar{N}'_{Ri} \, (y'_\nu)_{ij} \,
\tilde{H}^\dagger \, \ell'_{Lj} + h.c. \, \,, 
\eea 
here we have
defined $N_{Ri}^c = C \bar{N}_{Ri}^T$ with $C$ the charge conjugation
matrix.  There is freedom to rotate to the mass basis by introducing
the unitary rotation matricies $\mathcal{U}(f,L/R)$, so that 
\bea
N'_{Ri} &=& \mathcal{U}(N,R)_{ij} \, N_{Rj}, \quad \quad \ell'_{Li} =
\mathcal{U}(L,L)_{ij} \, \ell_{Lj}, \quad \quad e'_{Ri} =
\mathcal{U}(e,R)_{ij} \, e_{Rj}.  
\eea 
The kinetic terms are unchanged
by these rotations, and one is free to further rotate to a basis in
the flavour space of $M, \, y_E, \, y_\nu$ defined by the set of all
possible $M, \, y_E, \, y_\nu$ related through 
\bea
\label{Eq:y-rots}
M' &\rightarrow&  \mathcal{U}(N,R)^\star \, M \,  \mathcal{U}(N,R)^\dagger, \nn \\
y'_E &\rightarrow&  \mathcal{U}(L,L) \, y_E \,  \mathcal{U}(e,R)^\dagger,  \\
y'_\nu &\rightarrow& \mathcal{U}(N,R) \, y_{\nu} \,
\mathcal{U}(L,L)^\dagger. \nn 
\eea 
We choose to make the initial
rotation to a basis where $M$ is diagonal real and nonnegative. This
fixes $\mathcal{U}(N,R)$. Initially the Majorana mass matrix has three
complex eigenvalues $M_{a,b,c} = \eta_{a,b,c} \, |M_{a,b,c}| =
\eta_{a,b,c} \, (\mu_{a,b,c})$, here $\eta_{a,b,c}$ are the Majorana
phases. Working in this basis
\cite{Broncano:2002rw,Broncano:2003fq,Jenkins:2008rb} shifts any
Majorana phases into the $y_\nu$ matrix and $ N_{i} = \sqrt{\eta_i} \,
N_{Ri} + \sqrt{\eta_i^\star} \, N_{Ri}^c$.  Integrating out the heavy
$N_{i}$ one obtains the dimension five operator \cite{Weinberg:1979sa}
\bea 
\mathcal{L}_5 = \frac{1}{2} \, (\tilde{H}^\dagger \ell_i) \,
C_{ij} \, (\tilde{H}^\dagger \, \ell_j) + h.c. \, \, , 
\eea 
with the matrix Wilson coefficient $C_{ij} = (y_\nu^T \, \eta M^{-1} \, y_\nu)_{ij}$. The key observation that we use in this work is that
when integrating out the $N_{i}$ in sequence\footnote{We distinguish
  in this paper right handed neutrinos that are in a mass diagonal
  basis, reserving the notation $N_{i}$ for such states, compared to
  $N_{Ri}$ states, which are not mass diagonal in general.}  a
perturbative expansion for neutrino phenomenology that is related to a
hierarchy in the magnitude of the contributions to $C_{ij}$ can be
constructed. Note that this is also the key point in the sequential dominance
idea of Refs. \cite{King:1998jw,King:1999cm,King:1999mb,King:2002nf}, however, the formalism we will develop is distinct from 
these past results.

\subsection{Perturbing the Seesaw}

\subsubsection{Model Dependence of Perturbing the Seesaw}
The FSE we develop depends on the mass spectrum and the Yukawa
couplings of the $N_{i}$. Also, in principle, the flavour
orientation\footnote{By the flavour orientation of the Yukawa
  couplings, we mean the orientation of the Yukawa coupling vectors
  $\vec{x},\vec{y},\vec{z}$ with respect to the leading order
  eigenvectors $\vec \rho_{a,b,c}$; see the next section for definitions
  and further details.} effects the quality of the FSE.
Experimentally, at this time, all of these Lagrangian parameters are
individually unknown, thus we will be forced to assume that a
perturbative expansion of the form we employ exists. Our approach is
to view the seesaw Lagrangian given in Eq. (6) as an effective theory,
and we do not attempt to uniquely identify and restrict ourselves to a
particular UV physics that dictates the low energy parameters in this
effective theory in this paper.  However, the formalism we develop
allows perturbative investigations of the neutrino mass spectrum and
mixing angles in many scenarios and is in fact quite general. It would
be surprising if all of the unknown parameters conspired to forbid an
expansion of this form from being present. We seek to illustrate this
point in this section, by demonstrating a simple scenario where the
FSE would be of some utility.

There are of course cases when the formalism we outline cannot be used.
When the $\mu_{a,b,c}$ and Yukawa couplings of the  $N_{i}$ are both highly degenerate a 
FSE cannot be used.\footnote{An exception to this statement is when the flavour orientation is such that an expansion is present due to the geometry in flavour space, we discuss an example of this form in Section \ref{expan}.}
This is also the case if a non-degenerate pattern of the $\mu_{a,b,c}$ and the Yukawa couplings are such that the perturbations to the neutrino masses are
similar in magnitude. However, as the $\mu_{a,b,c}$ depend on the matching of the effective theory to UV physics, while the Yukawa
couplings are dimension four and are not UV sensitive, this would require tuning the physics of different energy scales.
Unless model building is used to relate the masses and Yukawa couplings of the $N_{i}$ 
so that the contributions to the $\nu_L$ masses are the same,
a FSE is expected quite generally, and can be related to neutrino mass differences, as we will show. 

One can study the expected spectrum of left handed masses due to Eqn. (\ref{seesaw}) in generic scenarios.
Without assuming other interactions beyond the SM, the $N'_{Ri}$ are not distinguished
by any quantum number. As a result, the non-diagonal  mass matrix that is the coefficient of the operator matrix $\mathcal{O}_{ij} = N^{'}_{Ri} \, N'_{Rj}$ given by $M'_{ij}$
is naively expected to have entries that are all similar in magnitude.  As the operator is dimension three, the mass matrix is expected to be proportional to the
highest scale UV physics (violating lepton number) that was integrated out leading to this effective theory. We denote this scale by $M_0$ and the naive non-diagonalized mass matrix in this case as
\bea\label{massmatrix}
M'_{ij} \simeq M_0 \,   \left( \begin{array}{ccc}
1 & 1 & 1 \\
1 & 1 & 1  \\
1 & 1 & 1 
\end{array} \right),
\eea
with eigenvalues $\mu_i^0 = \{3 \, M_0, 0,0\}$. The vanishing of the
two eigenvalues of $M'_{ij}$ can be lifted by interactions of the
$N'_{Ri}$. These interactions can be dictated by beyond the SM quantum
numbers assigned in UV model building.  If the mass matrix is still
approximately degenerate, as in Eq.(\ref{massmatrix}), then a
hierarchy of the $N_{i}$ masses is still expected.  An example of a
small breaking where other interactions do not need to be assumed is
given by the orientations of the $N'_{Ri}$ in flavour space. Rotating
to the lepton diagonal mass basis, these interactions give loop
corrections 
\bea 
\epsilon_{ij} =\delta M'_{ij}/M'_{ij} \simeq
\frac{(y^\prime_\nu)^*_{ik}(y^\prime_\nu)_{jk}}{16 \, \pi^2} \, \log
\left(\frac{\mu^2}{M_0^2}\right)\qquad\text{no sum on $i,j$}.  
\eea
Logarithmic corrections of this form are required to cancel the
renormalization scale dependence of the pole masses of the $N'_{Ri}$.
Including such corrections leads to $M'_{ij} \simeq M_0 \, (1 +
\epsilon_{ij})$. These corrections split the mass spectrum;
diagonalizing one finds
\bea
\mu_1 = M_0 \left(3 + \frac{\sum_{ij} \, \epsilon_{ij}}{3} \right), \quad \mu_2 = \frac{2}{3} \, M_0 \left[ \sum_{i} \, \epsilon_{ii} -\sum_{i < j} \, \epsilon_{ij} \right],  \quad \mu_3 =  M_0 \, \mathcal{O}(\epsilon^2). 
\eea 
A hierarchical spectrum of $\mu_i$ is expected with the pattern $(\mu_1, \mu_2 , \mu_3) \simeq M_0 (1,\epsilon, \epsilon^2)$.

In the case of degenerate $\mu_i$ due to the $M'_{ij}$ mass matrix not conforming to these generic expectations, the expansion can follow from a hierarchy in the Yukawa couplings of the $N_{i}$ -- such as
the hierarchical pattern of Yukawa couplings in the SM. Lastly, the suitability of the expansion can follow only from the orientation in flavour space of the Yukawa coupling vectors of the $N_i$.
In summary, the appropriateness of the FSE is clearly model dependent, but we expect it to be broadly applicable in realistic models.

\subsection{Developing the Seesaw Perturbations}\label{exp}

Consider supplementing the SM field content by a single right handed
neutrino\footnote{The subscript in $N_i$ will run over labels $a,b,c$;
  for now we are just considering the single spinor field $N_a$.}
$N_{a}$ and an interaction term that couples it into a linear
combination of $\ell_L^\beta$.  The coupling is fixed by the complex
Yukawa vector $\vec{x}^T = \{x_1, x_2, x_3 \}$ in flavour space whose
form can be constrained by a flavour symmetry, but is here left
arbitrary. We absorb the overall Majorana phase into this vector.  The
relevant Lagrangian terms\footnote{We have chosen $\mathcal{U}(N,R)$
  as discussed in Section \ref{seesaw} to rotate to a mass diagonal
  basis for $N_{i}$ and absorbed this rotation into a redefinition of
  the Yukawa matrices $y_\nu$. The Yukawa vectors $\vec{x}$, $\vec{y}$
  and $\vec{z}$ introduced below are given in this basis.} are given
by
\bea
\label{flavourcoup}
\mathcal{L}_a & = & - \frac{\mu_a}{2} N_{a} \, N_{a} - x_{\beta} \,
\bar{N_{a}} \, \tilde{H}^\dagger \, \ell_L^{\beta} + h.c.  \eea
The right handed neutrino can be integrated out, giving for the left
handed neutrino mass matrix a nonzero eigenvalue
\bea 
\mathcal{M}\equiv \frac{v^2}{2 \, \mu_a} {\vec{x} \, \vec{x}^T} =
{\cal{U}}(\nu,L)^\star\, {\rm diag}(0, 0, m_a)\,
\mathcal{U}(\nu,L)^\dagger.  
\eea
The matrix ${\cal{U}}(\nu,L)$ is the matrix
$(\vec{\rho_c}^\star,\vec{\rho_b}^\star,\vec{\rho_a}^\star)$ of normalized
(column) vectors ${\vec{\rho}_{a,b,c}}^{\, \star}$ that solve $\mathcal{M}\vec{\rho}^{\, \star}=
m\vec\rho$, with $m$ real and non-negative.\footnote{Here we use the notation that the eigenvectors run over the flavour indicies $a,b,c$ to
denote that they are associated with integrating out
each of the $N_{a,b,c}$ at leading order in the corresponding mass eigenvalues.} These vectors are also
eigenvectors of $\mathcal{M}^\dagger
\,\mathcal{M} = (\vec x^\star \vec x^\dagger) \, (\vec x \, \vec
x^T)=(\vec x^\dagger \vec x) \vec x^\star \, \vec x^T$ with eigenvalues $m^2$; their complex conjugates,
$\vec \rho$ are eigenvectors of $\mathcal{M} \,\mathcal{M}^\dagger = (\vec x \vec
x^T)\, (\vec x^\star \, \vec x^\dagger )=(\vec x^\dagger \vec x) \vec
x\vec x^\dagger$, also with
eigenvalues $m^2$. One finds the leading eigenvector and eigenvalue
\bea 
\vec{\rho_a} = {\vec{x}}/{|{\vec{x}}|}, \quad \quad m_a = v^2
({\vec x}^\dagger \vec{x})/2 \mu_a.  \eea
Here $\mu_a$ is real and non-negative due to the initial flavour basis
choice that fixed $\mathcal{U}(N,R)$, while $\mathcal{U}(\nu,L)$ and
${\vec{x}}$ are in general complex.\footnote{This choice is allowed by
  the unitary flavour transformations that are symmetries of the
  kinetic terms, Eqs.~(6--8).}  The remaining two eigenvectors
$\vec{\rho}_{b,c}$ are such that
$\langle\vec{\rho_a}|\vec{\rho_b}\rangle\equiv
\vec{\rho_a}^\dagger\vec{\rho_b} = 0$, $\langle\vec{\rho_a}
|\vec{\rho_c}\rangle = 0$.
These eigenvectors will lead to the remaining (smaller) mass
eigenvalues, and will perturb
the leading eigenvalue and eigenvector and thus the ${\cal{U}}(\nu,L)$
matrix. This is the expansion we seek to exploit. Consider the
perturbation that generates the second eigenvalue of the neutrino mass
matrix due to a second right handed neutrino $N_{b}$, which we define
to couple into a linear combination of the $\ell_L$ given by $\vec{y}^T =
\{y_1, y_2, y_3 \}$. One obtains a second eigenvalue in the left
handed neutrino mass matrix so long as $\vec{y} \not\parallel
\vec{x}$.  The perturbation of  $\mathcal{M} \,\mathcal{M}^\dagger$ is given by
\bea
(\mathcal{M} + \delta \mathcal{M}) \,(\mathcal{M} + \delta \mathcal{M})^\dagger  &=&  \frac{v^4}{4 \mu_a^2} \,  \vec x \, \vec x^T\, (\vec x^\star \, \vec x^\dagger ) + \frac{v^4}{4 \mu_a \, \mu_b} (\vec{x} \, \vec{x}^T)\, (\vec{y}^\star \, \vec{y}^\dagger) + \frac{v^4}{4 \mu_a \, \mu_b} (\vec{y} \, \vec{y}^T)\, (\vec{x}^\star \, \vec{x}^\dagger), \nn \\
&+& \frac{v^4}{4 \mu_b^2} \,  \vec y \, \vec y^T\, (\vec y^\star \, \vec y^\dagger ). 
\eea
  At leading order the $\vec{\rho}_{b,c}$ have degenerate
(vanishing) eigenvalues. One is free to rotate to a chosen basis in these vectors. We rotate to a basis in these vectors 
such that the following perturbation vanishes for $\vec{\rho}_{c}$
\bea
\langle \vec{\rho}_{c}| \,  (y y^\dagger)\, (y^T\, y^\star) \, | \vec{\rho}_{c} \rangle = 0.
\eea
With this choice $\vec{\rho}_c$ retains a vanishing eigenvalue when $N_{b}$ is integrated out.  The
eigenvector $\vec{\rho}_c$ should be orthogonal to $\vec{\rho}_a
\propto \vec{x}$. A normalized eigenvector basis at leading
order is then 
\bea 
\vec{\rho}_b = { \frac{\vec{x}^\star \times
    (\vec{y} \times \vec{x})}{|\vec{x}| \, | \vec{x} \times
    \vec{y}|}}, \quad \quad \quad \vec{\rho}_{c} = \frac{\vec{y}^\star
  \times \vec{x}^\star}{|\vec{x} \times \vec{y}|}.  
\eea 
Now we can determine the perturbation on the leading order
eigenvectors and eigenvalues when $N_{b}$ is integrated
out. The corrections to the eigenvalues and eigenvectors using perturbation theory are given by
\bea
\delta \vec{\rho_j} &=& = \sum_{i \ne j} \frac{\langle \vec{\rho}_i | \mathcal{M} \delta \mathcal{M}^\dagger   + \delta \mathcal{M} \, \mathcal{M}^\dagger | \vec{\rho}_j \rangle}{m_j^2 - m_i^2} \, \vec{\rho}_i, \nn\\
\delta m_i^2 &=&  \langle \vec{\rho}_i |\mathcal{M} \delta \mathcal{M}^\dagger   + \delta \mathcal{M} \, \mathcal{M}^\dagger +  \delta \mathcal{M} \,  \delta \mathcal{M}^\dagger   | \vec{\rho}_i \rangle. 
\eea  
Nonzero eigenvalues are obtained for $m_{b,c}$ at second order in the expansion due to the orthogonal basis vectors causing the leading perturbations to each of these masses to vanish. The leading perturbation to the eigenvectors
and $m_{a}$ is first order in the expansion. We retain the leading perturbation on the eigenvectors and the leading and subleading effects on the masses to obtain nonzero eigenvalues. We find the following for the perturbations
\bea
\delta  \vec{\rho_a} &=&  \mu^2_{ab}  \frac{\langle \vec{\rho_b}| \vec{y} \rangle (\vec y \cdot \vec x^\star) |\vec x|}{\delta m_{ab}^2} \, \vec{\rho_b},  \quad \quad  \quad \, \, \delta m_a^2 =  2 \,\mu^2_{ab}  \, {\rm Re}\left[\langle \vec{y}| \vec{x} \rangle \, (\vec{x}\cdot \vec{y}^\star) \right] + \mu^2_{bb} \, | \vec y|^4 \, \cos^2 \theta_{xy}, \nn \\
\delta  \vec{\rho_b} &=&  \mu^2_{ab} \frac{\langle \vec{y} | \vec{\rho_b} \rangle (\vec x \cdot \vec y^\star) |\vec x|}{\delta m_{ba}^2} \, \vec{\rho_a},  \quad \quad \quad \,  \, \delta m_b^2 =   \mu^2_{bb} |\langle \vec{\rho_b}| \vec{y} \rangle|^2 \, |\vec y|^2,  \, \nn \\
\delta \vec{\rho_c} &=& 0, \hspace{4.7cm} \delta m_c^2 = 0.  
\eea
Here $\delta m_{ij}^2 = m_i^2 - m_j^2$, $\mu^2_{ij} =v^4/(4 \, \mu_i
\, \mu_j)$ and $\cos \theta_{xy}=|\vec x^*\cdot \vec y|/|\vec x||\vec
y|$ is a measure of (the cosine of) the angle between
the $\vec x,\vec y$\ \;Yukawa vectors. 
The masses are evaluated to the appropriate
order in the perturbative expansion and the eigenvector perturbations
are in general complex. Note that for the phenomenology of the
$\mathcal{U}_{PMNS}$ matrix that we will study it will be sufficient
to only retain the leading perturbation, while when studying the mass
spectrum the leading and sub-leading perturbations should be retained.

Finally integrate out the third right handed neutrino $N_{c}$ with Majorana mass $\mu_c$, which
couples into a linear combination of the $\ell_L^\beta$ dictated by $z_\beta^T =\{z_1,z_2,z_3 \}$.  
The resulting eigenvector perturbations are
\bea
\delta_2 \vec{\rho_a} &=& \mu^2_{ac}\frac{\langle \vec{\rho_b}| \vec{z} \rangle (z \cdot x^\star) | \vec x|}{\delta m_{ab}^2} \,\vec{\rho_b}
+ \mu^2_{ac} \frac{\langle \vec{\rho_c}| \vec{z} \rangle (z \cdot x^\star) | \vec x|}{\delta m_{ac}^2} \,\vec{\rho_c},  \nn \\
\delta_2 \vec{\rho_b} &=& \mu^2_{ac}\frac{\langle \vec{z} | \vec{\rho_b} \rangle (x \cdot z^\star) | \vec x|}{\delta m_{ba}^2} \,\vec{\rho_a}, \nn \\
\delta_2 \vec{\rho_c} &=& \mu^2_{ac}\frac{\langle \vec{z} | \vec{\rho_c} \rangle (x \cdot z^\star) | \vec x|}{\delta m_{ca}^2} \,\vec{\rho_a}. 
\eea
The mass perturbations are
\bea
\delta_2 m_a^2 &=&  2 \,\mu^2_{ac}  \, {\rm Re}\left[\langle \vec{z}| \vec{x} \rangle \, (\vec{x}\cdot \vec{z}^\star) \right] + \mu^2_{cc} \, | \vec z|^4 \, \cos^2 \theta_{xz}, \nn \\
\delta_2 m_b^2 &=&  \mu^2_{cc} \,| \langle \vec{\rho_b} | \vec{z} \rangle |^2 \, |\vec z|^2 , \nn \\
\delta_2 m_c^2 &=&  \mu^2_{cc}\, | \langle \vec{\rho_c} | \vec{z} \rangle |^2 \, |\vec z|^2.
\eea
The measured masses of the neutrino's are related to these perturbations as
\bea
m_A^2 &=& m_a^2 + \delta m_a^2 + \delta_2 m_a^2, \nn \\
m_B^2 &=& \delta m_b^2 + \delta_2 m_b^2, \nn \\
m_C^2 &=& \delta_2 m_c^2. 
\eea
It is interesting to note that a normal hierarchy emerges quite naturally from the FSE as the leading neutrino mass $m_a$ receives corrections at linear order
to its mass, while the remaining masses only receive corrections at second order in the perturbations.

It is also important to note that this formalism does not require a hierarchy of the form $\delta_2 m_i^2 \ll \delta m_i^2$, only $\delta_2 m_i^2 , \delta m_i^2 \ll m_a^2$ is required.
Expanding on this important point in more detail, it is not required that the perturbation due to integrating out $N_b$ is larger than the perturbation due to integrating out $N_c$. Only
that the effect of integrating out each of these right handed neutrinos perturbs the initial mass matrix ---which is dominated by integrating out the initial right handed neutrino $N_a$.
The existence of these perturbations are not necessarily direct statements on the relative size of the $\mu_i$ as we discuss in more detail in the next section.  

\subsection{Inverted and Normal Hierarchy and Flavour Space}\label{expan}
For a normal hierarchy (with notation $m_{3} > m_{2} \gtrsim m_1$) we identify $(1,2,3) = (C,B,A)$ in the equations above.
The difference between the normal and inverted ($m_{3} < m_1 \lesssim m_2$) case appears in the relative size of the $\delta m_i$ and $\delta_2 m_i$ and the identification
$(C,B,A) = (3,1,2)$ for an inverted hierarchy. An inverted hierarchy requires non generic perturbations in the FSE, or
one can trivially modify the expansion to only perturb when $N_{c}$ is integrated out. The size of $\delta m_i$ and  $\delta_2 m_i$ depends on the hierarchy
in $\mu_{a,b,c}$, the magnitude of the Yukawa vectors, and also the orientation in flavour space of the vectors ${\vec{x}, \, \vec{y}, \, \vec{z}}$. In this section, we will discuss the size of the perturbations of the FSE in light of neutrino mass data in
a scenario where the perturbations are dictated primarily by a hierarchy
in $\mu_{a,b,c}$. We will then discuss the case where the expansion arises primarily due to the orientation of the Yukawa coupling vectors in flavour space.

\subsubsection{The expansion without flavour alignment}

We can examine the quality of the expansion by comparing to the measured mass differences in neutrinos.
Consider the case that the size of the perturbations is generic in the sense that the Yukawa coupling vectors are taken to be $\mathcal{O}(1)$
with the orientation in flavour space not significantly effecting the quality of the expansion. In this case, the  expansion follows from the relative size of the $\mu_i$, i.e. $\mu_a < \mu_b < \mu_c$.
Consider the generic case in a normal hierarchy. Using the FSE and retaining the dominant term
\bea
\Delta m_{32}^2 &=& m_a^2 + \delta m_a^2 + \delta_2 m_a^2 - (\delta m_b^2 + \delta_2 m_b^2) \approx  m_a^2,
\eea
while in the same manner $\delta m_b^2 \sim \Delta m_{21}^2$ and the expansion requires
$\delta_2 m_c^2 < \Delta m_{32}^2, \Delta m_{21}^2$. Due to this, the expansion requires
$v^2 |\vec{z}|^2/(2 \, \mu_c) < \sqrt{\Delta m_{32}^2} \sim 0.05 \, {\rm eV}$.
This condition is consistent with current bounds on the absolute neutrino mass scale,
with a $95 \%$ C.L. bound of $\sum m_\nu = 0.28 \, {\rm eV}$ quoted in Ref. \cite{Thomas:2009ae}, assuming $\Lambda{\rm CDM}$ cosmology. It is also consistent with current bounds
from Tritium $\beta$ decay experiments \cite{Otten:2008zz} which quote $m(\nu_e) < 2 \,  {\rm eV}$ at $95 \%$ C.L. 

Expressing this condition in terms of the high mass scale of the $N_c$ integrated out, $\mu_c/|\vec{z}|^2 \gtrsim 10^{14} \, {\rm GeV}$. 
Generically one expects the mass scale of the  right handed neutrino
operator to be the largest scale integrated out that violated $L$ number, and this condition for the lightest neutrino is clearly 
consistent with naive expectations of $M_0 \sim M_{pl}$. 

\subsubsection{The expansion with flavour alignment}

Now consider the case where the  perturbative expansion follows from the flavour orientation of the ${\vec{x}, \, \vec{y}, \, \vec{z}}$ vectors primarily.
An example where this is the case is when the  threshold matching onto the UV physics is such that the Wilson coefficient matrix of 
 $\mathcal{O}_{ij}$ yields a mass matrix with nearly degenerate eigenvalues. This occurs for example when
\bea
M'_{ij} \simeq M_0 \,   \left( \begin{array}{ccc}
1 + \epsilon & \epsilon & \epsilon \\
\epsilon & 1 + \epsilon & \epsilon  \\
\epsilon & \epsilon & 1 + \epsilon
\end{array} \right).
\eea
In this case, a nearly degenerate mass spectrum of the $N_R$ follows, $\mu_a \simeq \mu_b \simeq \mu_c$.
Generating a normal or inverted hierarchy if one also has $|\vec x| \sim  |\vec y| \sim |\vec z|$ requires more precise alignments in flavour space and allows a geometric interpretation of the
measured neutrino mass spectrum. In the FSE, the tree level masses of the SM neutrinos are
\bea
m_A^2 &=& \mu_{aa}^2 \,  |\vec{x}|^4 + 2 \mu_{ab}^2 |\vec{x}|^2 |\vec y|^2 \cos^2 \theta_{xy} + 2 \mu_{ac}^2 |\vec{x}|^2 |\vec z|^2 \cos^2 \theta_{xz} + \mu_{bb}^2  |\vec{y}|^4 \, \cos^2 \theta_{xy}  + \mu_{cc}^2  |\vec{z}|^4 \, \cos^2 \theta_{xz}, \nn \\
m_B^2 &=& \mu_{bb}^2  |\vec{y}|^4 \, \cos^2 \theta_{\rho_b y}  + \mu_{cc}^2  |\vec{z}|^4 \, \cos^2 \theta_{\rho_b  z}, \nn \\
m_C^2 &=& \mu_{cc}^2  |\vec{z}|^4 \, \cos^2 \theta_{ \rho_c  z}.
\eea
Consider the case that all of the $\mu_{ij}^2$ are similar in magnitude and $|\vec x| \gtrsim |\vec y| \sim |\vec z|$ so that the mass spectrum is primarily dictated by the orientation of the Yukawa vectors in flavour space.
An example of an inverted or normal hierarchy is shown in Fig. (1) in this case.
\begin{figure}[htbp]
\centerline{
\includegraphics[width=0.8\textwidth]{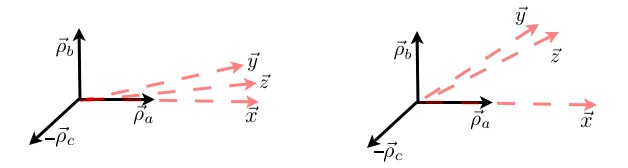}}
\caption{A geometric interpretation of an inverted or normal hierarchy when  $\mu_a \simeq \mu_b \simeq \mu_c$ and ${|\vec{x}| \gtrsim |\vec{y}| \sim |\vec{z}|}$.
Normal hierarchy on the left, inverted on the right.}
\end{figure}

\subsection{$\mathcal{U}_{PMNS}$ and Flavour Space}

Now, assuming that a FSE expansion exists, let us consider its utility in examining the form of $\mathcal{U}_{PMNS}$.
The rotation matrix  $\mathcal{U}(\nu,L)$ with $\vec{v}_i^\star =  (\vec{\rho}_i + \delta  \vec{\rho_i} + \delta_2  \vec{\rho_i})^\star$ is given by $\mathcal{U}(\nu,L)  = \left(\vec{v}_3^\star  , \vec{v}_2^\star ,  \vec{v}_1^\star \right)$.
The charged lepton mass matrix after electroweak symmetry breaking is given by $\mathcal{M}_e = v \, y_E/\sqrt{2}$ and diagonalized by\footnote{The hierarchy of the charged lepton masses can be used to organize an expansion of $\mathcal{U}(e,L)$ in the same manner by constructing
$\mathcal{M}_e^\dagger \mathcal{M}_e$ in principle. Conversely, in principle, one could employ an ansatz that the hierarchy in these mass eigenvalues could be related to
an expansion of $\mathcal{U}(e,R)$. As such we do not employ a FSE on $\mathcal{U}(e,L)^\dagger$.
This ambiguity also limits the application of a FSE to the $\rm CKM$ matrix.
In this manner, the expansion we employ is most useful for expanding $ \mathcal{U}(\nu,L)$ when a seesaw mechanism is the origin of the smallness of the neutrino masses.} 
\bea
\mathcal{U}(e,R) \, \mathcal{M}_e \, \mathcal{U}^\dagger(e,L) = {\rm diag}(m_e, m_\mu,m_\tau).
\eea
We take $\mathcal{U}(e,L)^\dagger  =  \left( \vec{\sigma_3}^\dagger , \vec{\sigma_2}^\dagger , \vec{\sigma_1}^\dagger \right)$,
with $\vec{\sigma}_i$ the orthonormal column eigenvectors diagonalizing $\mathcal{M}_e \mathcal{M}_e^\dagger$.
The expanded $\mathcal{U}_{PMNS}^\star$ is of the form
\bea\label{UPMNS}
\mathcal{U}_{PMNS}^\star &=& \left( \begin{array}{ccc}
\vec{v}_3 \cdot  \vec{\sigma}_3 & \vec{v}_2 \cdot  \vec{\sigma}_3 & \vec{v}_1 \cdot  \vec{\sigma}_3  \\
\vec{v}_3 \cdot \vec{\sigma}_2  &\vec{v}_2 \cdot  \vec{\sigma}_2  &\vec{v}_1 \cdot  \vec{\sigma}_2   \\
\vec{v}_3 \cdot  \vec{\sigma}_1 & \vec{v}_2 \cdot  \vec{\sigma}_1   &\vec{v}_1 \cdot  \vec{\sigma}_1 \end{array} \right), \nn \\
&=& \left( \begin{array}{ccc}
\vec{\rho}_3 \cdot  \vec{\sigma}_3 & \vec{\rho}_2 \cdot  \vec{\sigma}_3 & \vec{\rho}_1 \cdot \vec{\sigma}_3  \\
\vec{\rho}_3 \cdot  \vec{\sigma}_2  & \vec{\rho}_2 \cdot  \vec{\sigma}_2  & \vec{\rho}_1 \cdot \vec{\sigma}_2   \\
\vec{\rho}_3 \cdot \vec{\sigma}_1 & \vec{\rho}_2 \cdot  \vec{\sigma}_1   & \vec{\rho}_1 \cdot  \vec{\sigma}_1 \end{array} \right) + \mathcal{O}( \delta \vec{\rho}_i, \delta_2 \vec{\rho}_i).
\eea
This result makes clear that the first two right handed neutrinos that were integrated out in the FSE
contribute to the leading order structure of the PMNS matrix. The
leading order of the $\vec \rho_i$ eigenvectors
only depended on ${\vec{x},\vec{y}}$. As we have discussed, these neutrinos can be integrated
out in sequence, and for Yukawa couplings that are degenerate the {\it lightest} two $N_i$ can lead to the largest mass eigenvalues of the $\nu_L$.
In a mass degenerate case, they are more strongly coupled to the $\ell_{Lj}$. The unitarity of the $\mathcal{U}_{PMNS}$ matrix is ensured when an expansion of this form is employed  by normalizing the $\vec{v}_i$
eigenvectors order by order in the expansion. Conversely, if the FSE is used when the expansion parameters are not small, the constructed $\mathcal{U}_{PMNS}$ matrix is not necessarily unitary.

It is important to note that, in general, it is the relationship between the eigenvectors that determines $\mathcal{U}_{PMNS}$. Exact $\mathcal{TB}$ form at leading order in the FSE has a simple interpretation.
It follows from the relationship between the 
eigenvector $\vec{\rho}_1$ and the $\sigma_i$ being $\vec{\rho}_1 = (\vec{\sigma}_1 - \vec{\sigma}_2)/\sqrt{2}$ in our chosen phase convention.

\section{(Un)Relating Flavour Symmetries to the $U_{PMNS}$ form}

This formalism can be used to systematically explore symmetries that are consistent with ${TB}$ form. Consider an exact $(\mathcal TB)$ form of $\mathcal{U}_{PMNS}$ at leading order in the FSE. 
This is dictated by $ \theta_{13} \equiv 0$ and $\theta_{23}$ maximal. We restrict the Yukawa vectors in this discussion to real values for simplicity.
Including an unfixed $s_{12}$ this gives (in a particular phase convention)
\bea
\mathcal{U}_{TB} = \left( \begin{array}{ccc}
c_{12} & s_{12} & 0 \\
-\frac{s_{12}}{\sqrt{2}} &  \frac{c_{12}}{\sqrt{2}}  & - \frac{1}{\sqrt{2}}  \\
-\frac{s_{12}}{\sqrt{2}} &  \frac{c_{12}}{\sqrt{2}}  & \frac{1}{\sqrt{2}}  \end{array} \right).
\eea
Assuming $\mathcal{U}_{PMNS} = \mathcal{U}_{TB}$ 
\bea\label{relations1}
\vec{\rho}_1 &=& \frac{1}{\sqrt{2}} \, \left(\vec{\sigma}_1 - \vec{\sigma}_2 \right),  \quad \quad \quad \quad \quad \quad \quad \vec{\sigma}_1 = \frac{1}{\sqrt{2}} (\vec{\rho}_1 + c_{12} \, \vec{\rho}_2  - s_{12} \, \vec{\rho}_3), \nn \\
\vec{\rho}_2 &=& \frac{c_{12}}{\sqrt{2}} (\vec{\sigma}_1 + \vec{\sigma}_2) + s_{12} \, \vec{\sigma}_3, \quad \quad \quad \quad \! \vec{\sigma}_2 = \frac{1}{\sqrt{2}} (- \vec{\rho}_1 + c_{12} \, \vec{\rho}_2  - s_{12} \, \vec{\rho}_3),\nn \\
\vec{\rho}_3 &=& -\frac{s_{12}}{\sqrt{2}} (\vec{\sigma}_1 + \vec{\sigma}_2) + c_{12} \, \vec{\sigma}_3, \quad \quad \quad \,  \vec{\sigma}_3 =  s_{12} \, \vec{\rho}_2  + c_{12} \, \vec{\rho}_3.
\eea

Consider the ``flavour" basis as an illustrative example of the FSE, where
$\vec{\sigma}_1 = (0,0,1), \, \vec{\sigma}_2 = (0,1,0), \, \vec{\sigma}_3 = (1,0,0)$. In this case,  $\mathcal{U}(e,L) = {\rm diag} (1,1,1)$ 
and $\mathcal{U}_{PMNS} =  \mathcal{U}(\nu,L)$. 
The flavour basis is appealing in that it allows the solution for the perturbations to the $\vec{\rho}_i$ eigenvectors
and the $\mathcal{U}_{PMNS}$ matrix  to proceed simply through solving
\bea
\frac{v^2}{2 \, \mu_a}  { {\vec{x} \, \vec{x}^T}} + \frac{v^2}{2 \, \mu_b}  { {\vec{y} \, \vec{y}^T}} &=&  {\cal{U}}_{TB}^\star \, {\rm diag}(0, \delta m_b, m_a + \delta m_a )\,  \mathcal{U}_{TB}^\dagger, \\
{\vec{x} \, \vec{x}^T} + \frac{\mu_a}{\mu_b} {\vec{y} \, \vec{y}^T} &=& \frac{(m_a + \delta m_a) \, \mu_a}{v^2} \, \left( \begin{array}{ccc}
0 & 0 & 0  \\
0  & 1  & -1   \\
0 & -1  & 1  \end{array} \right) +  \frac{\sqrt{2} \, \delta m_b \, \mu_a}{v^2} \left( \begin{array}{ccc}
 \sqrt{2} \, s_{12}^2 & s_{12} \, c_{12} &s_{12} \, c_{12}  \\
s_{12} \, c_{12}  & \frac{c_{12}^2}{\sqrt{2}}  & \frac{c_{12}^2}{\sqrt{2}}   \\
s_{12} \, c_{12} & \frac{c_{12}^2}{\sqrt{2}}  & \frac{c_{12}^2}{\sqrt{2}}  \end{array} \right). \nn
\eea 
Trivially one finds
\bea
\vec{x}^T =   \left(0,-1,1 \right) \sqrt{(m_a + \delta m_a) \, \mu_a}/v, \quad \quad \quad \vec{y}^T =   \left(\sqrt{2} \, s_{12},c_{12},c_{12} \right) \sqrt{\delta m_b \, \mu_b}/v.
\eea
It follows that in the flavour basis  $\vec{x} \cdot \vec{y} =0$ so  $ \delta m_a = 0$  and $ (\delta  \vec{\rho_a},  \delta  \vec{\rho_b}, \delta  \vec{\rho_c}) = (0,0,0)$.
Now consider including a third neutrino eigenvalue, retaining the required perturbation. For a solution, $z_2 = z_3$ is required, and consequently $\delta_2 m^2_a = 0$ as  $\vec{x} \cdot \vec{z} =0$.
As a result the leading eigenvector $\vec{\rho}_a$ is unperturbed by integrating out both $N_{b},N_{c}$ in the flavour basis, one finds 
\bea
\vec{z}^T = \pm \frac{\sqrt{2 \mu_c \, \delta_2 m_c}}{v} \left( \sqrt{c_{12}^2 + \frac{\delta_2 m_b}{\delta_2 m_c} \, s_{12}^2}, \frac{1}{\sqrt{2}} \, \sqrt{\frac{\delta_2 m_b}{\delta_2 m_c} \, c_{12}^2 + s_{12}^2} ,  \frac{1}{\sqrt{2}} \, \sqrt{\frac{\delta_2 m_b}{\delta_2 m_c} \, c_{12}^2 + s_{12}^2}\right).
\eea
A $\mu \leftrightarrow \tau$ symmetry implemented on $\vec{y}$ and $\vec{z}$ is consistent with $\mathcal{TB}$ form of the PMNS matrix, as expected.

\subsection{Perturbative breaking of $\mathcal{TB}$ form}
Now consider the breaking of $\mathcal{TB}$ form.
The value of  $\theta_{13}$ measured by the DAYA-Bay collaboration is in agreement with the global fit values given in Ref. \cite{Fogli:2011qn}, as such, we use the 
fit results to find the measured pattern of deviations in $\mathcal{TB}$ form. It is instructive to construct the following ratios of experimental values
characterizing the deviations of $\mathcal{TB}$ form in each mixing angle. Using the small angle approximation 
\bea
\frac{\tan^2(\delta \theta_{12})}{\sin^2(\delta \theta_{13})} =  0.02 \pm 0.32 \quad \quad  \frac{\tan^2(\delta \theta_{23})}{\sin^2(\delta \theta_{13})} =  0.01 \pm 0.10 \quad \quad \frac{\tan^2(\delta \theta_{12})}{\tan^2(\delta \theta_{23})} = 2.0 \pm 36.
\eea
Here we have used the one sigma new reactor flux values of Ref. \cite{Fogli:2011qn} and taken a $\pm$ symmetric one sigma error.
Various breaking of $\mathcal{TB}$ form can be studied using the FSE and compared to these results. Consider the case that $N_{b}$ retains a $\mu \leftrightarrow \tau$ flavour symmetry 
in its couplings to the charged leptons, but $N_{c}$ breaks such a symmetry so that deviations in $\mathcal{TB}$ form are expected. Fix 
$\vec{z}^{T'} = \vec{z}^T + (0,\Delta_1, \Delta_2)$ with $\Delta_1 \neq \Delta_2$ and treat this breaking as a perturbation using the FSE.
We  use $\Delta m^2_{AB} \simeq \Delta m^2_{AC}$ assuming a normal hierarchy. At leading order in the FSE the breaking of ${TB}$ has the pattern
\bea
\frac{\tan^2(\delta \theta_{12})}{\sin^2(\delta \theta_{13})} &=& 0, \nn \\
\frac{\tan^2(\delta \theta_{23})}{\sin^2(\delta \theta_{13})}  &=& \frac{2 \, \cos^2 \theta_{\rho_b z} +  \cos^2 \theta_{\rho_c z}}{ \cos^2 \theta_{\rho_b z}+ 2 \,   \cos^2 \theta_{\rho_c z}}, \nn \\
\frac{\tan^2(\delta \theta_{12})}{\tan^2(\delta \theta_{23})}  &=& 0.
\eea
As the range of the predicted value of $\tan^2(\delta \theta_{23})/\sin^2(\delta \theta_{13})$ is given by $[0.5,2]$ at leading order in the FSE, this pattern of flavour breaking does not reproduce the data.
In this way, particular mechanisms of the breaking of $\mathcal{TB}$ form can be falsified.
One can perform the exercise of assuming $\mathcal{TB}$ form is broken in the flavour basis by $N_{b}$ so that 
$\vec{y}^{T'} = \vec{y}^T + (0,\Delta_1, \Delta_2)$.  Using the expansion one finds the pattern of deviations are distinct. 
These breakings of $\mathcal{TB}$ form are also correlated with mass perturbations in each case which are trivial to determine using this formalism.

We emphasize however that the relationships {\it between} the eigenvectors determine the form of the PMNS matrix in general.
This is easy to demonstrate in more detail. Consider retaining a $\mu \leftrightarrow \tau$ symmetry imposed on $\vec{y}$ and $\vec{z}$ but deviating from the flavour basis,
choosing  $\vec{x}, \,  \vec{y}, \,  \vec{z}$ as above but unfixing $\vec{\sigma}_i$. At leading order in the expansion of $\mathcal{U}_{PMNS}$ one can solve for the condition that $\mathcal{TB}$ form
is still obtained for general $\vec{\sigma}_i$. One finds that {\it only} the flavour basis for $\vec{\sigma}_i$ gives a valid solution at leading order in the expansion.
This makes clear that $\mu \leftrightarrow \tau$ symmetry imposed on the  Lagrangian alone is not related to  $\mathcal{TB}$ form in a basis independent manner. 

As a further example that $\mu \leftrightarrow \tau$ symmetry is also not unique or of particular physical significance in allowing $\mathcal{TB}$ form (with an appropriate choice on the $\sigma_i$ eigenvectors),
consider the following procedure. Choose a $(e, \mu, \tau)$ symmetry on the first interaction eigenvector, ie  $\vec{x}^T = (1,1,1)/\sqrt{3}$, and  $\vec{y}^T = (1,1,0)/\sqrt{2}$ as a simple
interaction eigenvector for $N_b$ leading to an orthonormal eigenbasis at leading order, finding
\bea
\mathcal{U}(\nu,L) =  \left( \begin{array}{ccc}
\frac{1}{\sqrt{2}} &  \frac{1}{\sqrt{6}} & \frac{1}{\sqrt{3}}  \\
-\frac{1}{\sqrt{2}}  & \frac{1}{\sqrt{6}}  & \frac{1}{\sqrt{3}}   \\
0 & -\sqrt{\frac{2}{3}}  &\frac{1}{\sqrt{3}}  \end{array} \right).
\eea
Solving directly for $\mathcal{U}(\nu,L)$ so that at leading order $\mathcal{TB}$  form is obtained,
one finds
\bea
\mathcal{U}(e,L) = \left( \begin{array}{ccc}
\frac{1}{6}\left(\sqrt{2} + 2 \sqrt{3}\right)& \quad \quad \frac{1}{6}\left(\sqrt{2} - \sqrt{3} - \sqrt{6}\right) & \quad \quad  \frac{1}{6}\left(\sqrt{2} - \sqrt{3} - \sqrt{6}\right)   \\
\frac{1}{6}\left(\sqrt{2} - 2 \sqrt{3}\right) &  \quad \quad \frac{1}{6}\left(\sqrt{2} + \sqrt{3} - \sqrt{6}\right)  &  \quad \quad  \frac{1}{6}\left(\sqrt{2} + \sqrt{3} + \sqrt{6}\right) \\
-\frac{\sqrt{2}}{3} & -\frac{2 + \sqrt{3}}{3 \, \sqrt{2}}  & -\frac{\sqrt{2}}{3} + \frac{1}{\sqrt{6}} \end{array} \right).
\eea
This procedure can be used for any flavour symmetry chosen to fix $\vec{x},\vec{y}$ in the FSE for the $N_i$. 

We also note that  the impact of sterile neutrinos weakly coupled to the SM on neutrino phenomenology can be systematically studied with this approach.
For example, one can relate any measured value of a deviation from $\mathcal{TB}$ form to the particular Yukawa coupling vector of a single sterile neutrino, which can be 
shown to accommodate the value of $\theta_{13}$ reported by the Daya-Bay collaboration while the three right handed neutrinos partners of 
the SM fields give an exact $\mathcal{TB}$ form of the $\mathcal{U}_{PMNS}$ matrix.

\section{Conclusions}
Flavour symmetries that are related to the structure of the $\mathcal{U}_{PMNS}$ matrix only in a particular basis choice of $\mathcal{U}(e,L)$ 
can lead to suspect physical conclusions. As an alternative to basis dependent symmetry studies, we have developed a perturbative  expansion relating the measured 
masses of the neutrinos to the form of the PMNS matrix. This expansion offers a promising framework for broadly understanding the implications of the 
systematically improving experimental neutrino data, particularly in a normal hierarchy. 

We have illustrated our approach in an example where the flavour basis was chosen, for the sake of familiarity, and then shown how the expansion can control the predictions of $\mathcal{TB}$ form being broken. 
However, the approach we outline can accommodate any
basis choice. Indeed, it is the relationships between the eigenvectors that dictate the form of the  $\mathcal{U}_{PMNS}$ matrix in a  $\mathcal{U}(e,L)$ basis independent manner. 
This formalism can be employed in model building to attempt to determine a compelling origin of the eigenvector relationship that corresponds to $\mathcal{TB}$ form at leading order in the FSE. 
Further,  as the breaking of $\mathcal{TB}$ form is now experimentally established due to the discovery of a non zero $\theta_{13}$ in Ref. \cite{An:2012eh}, 
we expect the FSE to be of some phenomenological utility in falsifying mechanisms of the breaking of the
$\mathcal{TB}$ form of the $\mathcal{U}_{PMNS}$ matrix, as the pattern of this breaking is further resolved experimentally in the years ahead.

\newpage
\subsection*{Acknowledgments}
We thank  Mark Wise for collaboration in the initial stages of this work. We also thank Enrique Martinez for useful conversations and C. Grojean for comments on the manuscript.
The work of B.G. was supported in part by the US Department of Energy under contract DOE-FG03-97ER40546.
We thank the Aspen Centre for Theoretical Physics for hospitality.
 

\end{document}